\documentclass[aps, prb,twocolumn,showpacs,superscriptaddress]{revtex4-1}

\usepackage{graphicx}
\usepackage{mathrsfs}
\usepackage{dcolumn}
\usepackage{bm}
\usepackage{amsmath}
\usepackage{amsfonts}
\usepackage{color}
\usepackage[colorlinks,
            linkcolor=blue,
            anchorcolor=blue,
            citecolor=blue]{hyperref}

\begin{document}

\title{ Confinement effect enhanced Stoner ferromagnetic instability in monolayer 1T-VSe$_2$}

\author{Junyi He}
\author{Q. Xie}

\affiliation{Wuhan National High Magnetic Field Center and School of Physics, Huazhong University of Science and Technology, Wuhan 430074, China}
\author{Gang Xu}
\email{gangxu@hust.edu.cn}
\affiliation{Wuhan National High Magnetic Field Center and School of Physics, Huazhong University of Science and Technology, Wuhan 430074, China}

\date{\today}

\begin{abstract}


Monolayer 1T-VSe$_2$ has been reported as a room-temperature ferromagnet. 
In this work, by using first-principles calculations, 
we unveil that the ferromagnetism in monolayer 1T-VSe$_2$ is originated from 
its intrinsic huge Stoner instability enhanced by the confinement effect, 
which can eliminate the interlayer coupling, and lead to a drastic increase 
of the density of states at the Fermi level due to the presence of Van Hove singularity. 
Our calculations also demonstrate that the Stoner instability is very sensitive 
to the interlayer distance. 
These results provide a useful route to modulate the 
nonmagnetic to ferromagnetic transition in few-layers or bulk 1T-VSe$_2$, 
which also shed light on the enhancement of its Curie temperature 
by enlarging the interlayer distance.


\end{abstract}

\maketitle

\section{Introduction}

Ferromagnetic order in two-dimensional (2D) materials
is a highly desirable property that provides a new physical
degree of freedom to manipulate spin behaviors in spintronic
devices~\cite{Burch2018, Gibertini2020}. 
Previously, magnetism in 2D was mainly realized through depositing
films onto magnetic substrates, magnetic atoms adsorption, or 
doping~\cite{Gonzalez2016, Nair2012, Avsar2014}.
The shortcomings of these methods are obvious: (i)
one does not have an ideal 2D system from depositing and it is
impractical to integrate with spintronic devices, and (ii) disorder effects make the electronic properties
hard to design. Due to these drawbacks, 2D materials with intrinsic magnetic order have been actively pursued.

The CrI$_3$~\cite{Huang2017}
and Cr$_2$Ge$_2$Te$_6$~\cite{Gong2017}
are the first two experimentally reported
2D materials exhibiting long-range ferromagnetic (FM) order
with the Curie temperatures $T_c$ $\sim$ 45 K
and 30 K, respectively~\cite{Huang2017, Gong2017}.
These discoveries have stimulated
numerous research interests on
2D magnetic materials.
Very recently,
several materials with higher $T_c$ have been experimentally and theoretically explored,
including
MnSe$_x$~\cite{OHara2018},
Fe$_3$GeTe$_2$~\cite{deng2018gate},
and
1T-VSe$_2$~\cite{bonilla2018strong,
C3TC31233J,
doi:10.1002/adma.201903779,
doi:10.1021/acsnano.9b02996,
PhysRevB.101.035404}.
Among them, the 1T-VSe$_2$ is of particular interest 
since the bulk 1T-VSe$_2$ has a van der Waals (vdW) 
nature, which can be easily exfoliated 
to few-layers thickness. 
This gives 1T-VSe$_2$ the advantage  
to be tailored and manipulated for nano spintronic devices 
at low cost.


However, 
the nature of the ground phase of 1T-VSe$_2$ 
is still under hot debate.
Two groups have observed charge-density-wave (CDW) ground states and
concluded that
magnetic order is absent in the monolayer limit due to the CDW
suppression~\cite{doi:10.1021/acs.jpcc.9b08868,
doi:10.1021/acs.jpcc.9b04281}. 
Wong \emph{et al.} claimed
that a spin frustrated phase was observed
and the FM phase must be attributed to extrinsic 
factors~\cite{doi:10.1002/adma.201901185}.
Chua \emph{et al.}~\cite{doi:10.1002/adma.202000693} 
and Yu \emph{et al.}~\cite{doi:10.1002/adma.201903779} 
suggested
that the observed FM is \emph{not}
intrinsic, but caused by defects.
Nevertheless, Bonilla \emph{et al}.~\cite{bonilla2018strong} 
and many others~\cite{C3TC31233J,
doi:10.1002/adma.201903779,
doi:10.1021/acsnano.9b02996,
PhysRevB.101.035404} 
have presented strong experimental evidences 
for intrinsic 2D magnetism in monolayer 1T-VSe$_2$, 
which also reported a NM to FM phase transition 
from bulk to the monolayer limit~\cite{VANBRUGGEN1976251}. 
This is in contrast with
other 2D magnetic materials, 
where the FM phase is more stable in the bulk system.


\begin{figure}
\includegraphics[width=0.48\textwidth]{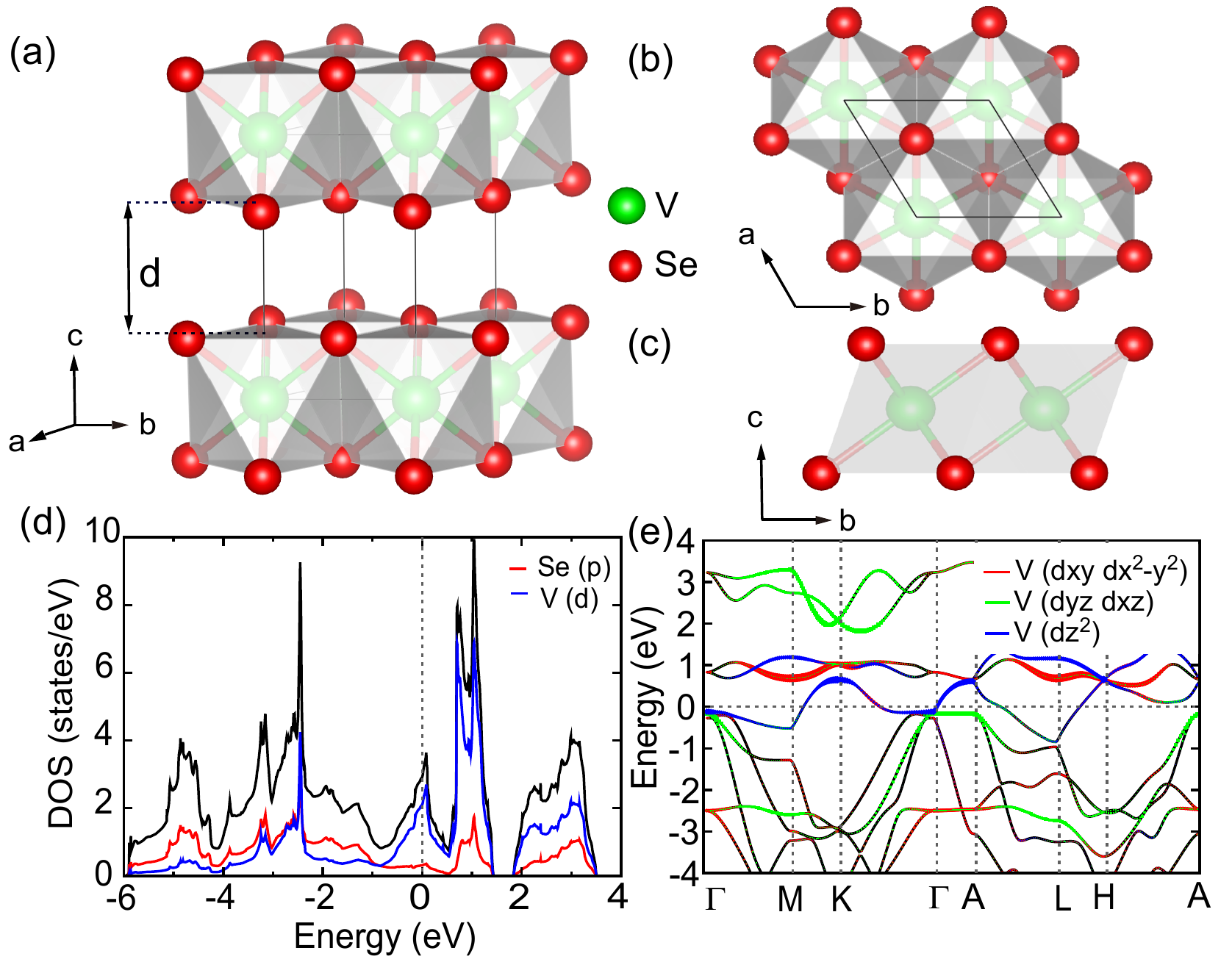}
    \caption{ (a) The crystal structure of bulk 1T-VSe$_2$. $d$ denotes the interlayer distance. 
    (b) and (c) are the top view and side view of the VSe$_2$ monolayer, respectively.
    (d) The total and projected DOS of bulk 1T-VSe$_2$ in the NM phase.
    (e) NM band structures of bulk VSe$_2$ with V-3$d$ orbitals projections.
}
\label{fig1}
\end{figure}

In this paper, we study the ground-state properties of 1T-VSe$_2$
by first-principles calculations. Through a comprehensive study
of the density of states (DOS) and band structures of
the bulk and few-layers 1T-VSe$_2$, we reveal that the monolayer
system has the strongest FM instability due to the presence of
Van Hove singularity (VHS) originated from saddle points at the Fermi level. 
We also find that 
in the few-layers case,
the couplings of $d_{z^2}$ orbitals between interlayer V atoms
split the saddle points away from the
Fermi level and weaken the FM instability.
The strongest FM instability in the monolayer is confirmed by the largest energy difference
between the NM and FM phases
and also verified by
using the phenomenological Stoner theory~\cite{stoner1938collective,
PhysRevB.16.255, PhysRevB.41.7028, PhysRevB.93.134407}. 
We thus conclude that the room-temperature 
FM order in the monolayer 1T-VSe$_2$ 
is intrinsic due to its unique electronic structures. 
Finally, we study the FM instability with respect to the interlayer distance $d$ [see Fig.~1(a)]
and predict that it is possible to 
tune the NM to FM phase transition in 
few-layers 1T-VSe$_2$ 
by enlarging the interlayer distance $d$. 
Our study 
provides an explanation 
to the origin of FM order in monolayer 1T-VSe$_2$ 
and also proposes a mechanism to tune 
the NM to FM phase transition 
in few-layers 1T-VSe$_2$. 


\begin{figure}
\includegraphics[width=0.48\textwidth]{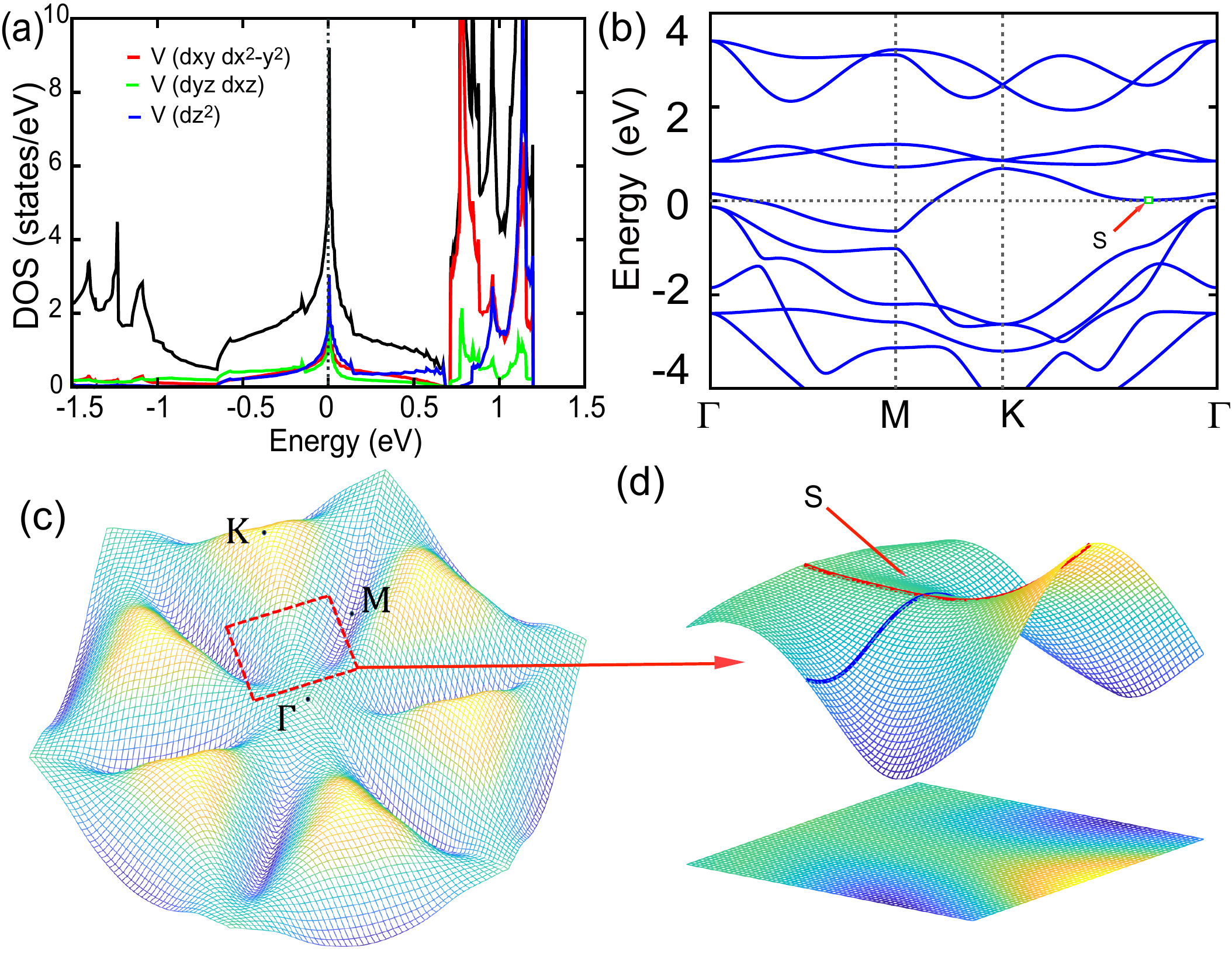}
\caption{
Results of the monolayer 1T-VSe$_2$ in the NM phase.
(a) The total DOS and projected DOS of the V-$3d$ orbitals.
(b) The corresponding band structures along the high symmetry $k$-path.
(c) Three dimensional plot of the $d_{z^2}$ band in the first BZ.
(d) Zoom-in band dispersion near the saddle point $S$.
}
\label{fig2}
\end{figure}

\section{Computational details }

VSe$_2$ usually adopts the 2H and 1T structures.
While the 2H-VSe$_2$ shows semiconducting behavior,
the 1T-VSe$_2$ is a metal~\cite{ma2012evidence, fuh2016metal,lebegue2013two}
and shows strong experimental evidence for FM ordering in the 
few-layers limit~\cite{bonilla2018strong}.
Different from the triangular prismatic crystal field in the 2H structure,
1T-VSe$_2$ has an octahedral crystal structure and belongs to $P\bar{3}m1$ space group,
where V atoms form a triangular lattice
and each V atom occupies the center of the octahedron surrounded by six Se atoms,
as shown in Figs. 1(a), 1(b) and 1(c).
As a result, each layer of VSe$_2$ is
stoichiometric~\cite{zhuang2016stability}.
The bulk crystal is composed of an AA stacking of VSe$_2$ sandwiches.

First-principles calculations based on density functional theory are
carried out by using the Vienna
\emph{ab initio} simulation package (VASP)~\cite{kresse1996efficient}.
The Perdew-Burke-Ernzerhof functional~\cite{perdew1996generalized}
is employed to treat the exchange-correlation interactions.
The cutoff energy for wave function expansion is set to 500 eV.
We use
19$\times$19$\times$9 and $21$$\times$21$\times$1 $\Gamma$-centered k meshes
to sample the Brillouin zone (BZ) in the bulk and slab calculations, respectively.
Structures are optimized until the force on each atom is less than 0.001~eV/{\AA}.
A vacuum layer of 15 {\AA} is set to minimize artificial interactions
between layers in the slab calculations.
For the bulk calculations, the
lattice constants $a=b=3.356$~{\AA},
and $c=6.105$~{\AA} are used~\cite{BAYARD1976325}.

\section{Results And Discussion}

We first calculate and plot the
total and projected DOS 
of the NM bulk 1T-VSe$_2$ 
in Fig.~1(a), which are in good agreement
with previous results~\cite{reshak2004theoretical, li2014versatile, feng2018electronic, yang2018metallic}.
The density at the Fermi level is about 2.9~states/eV, confirming its 
metallic nature. 
The projected DOS demonstrates 
that the states between $-0.9$~eV and $3.5$~eV
are mainly contributed by the V-$3d$ orbitals.
In an octahedron crystal field,
the five 3$d$ orbitals split into the lower $t_{2g}$
and the upper $e_g$ manifolds, 
which mainly contribute to the DOS 
around $-0.9\sim1.4$~eV and $1.9\sim 3.5$~eV, respectively. 
Furthermore,
due to the presence of a triangular field,
the $t_{2g}$ manifold
splits into
the lower $a_{1g}$ ($d_{z^2}$) and upper $e_g^\prime$
($d_{xy}$ and $d_{x^2-y^2}$) orbitals.
This picture is also verified from
the projected band structures 
in Fig.~1(e), 
in which the $d_{z^2}$
is the lowest 3$d$ orbital
that crosses the Fermi level and
dominates the low-energy physics of the bulk 1T-VSe$_2$.

We further plot the total and projected DOS
of the NM monolayer 1T-VSe$_2$ in Fig.~2(a). 
Compared with the bulk DOS in Fig.~1(d), 
we notice that these two DOS plots share many similarities. 
For instance, they both have high densities around 1~eV 
and they both possess energy gaps at $\sim$1.4~eV. 
This is reasonable due to that 1T-VSe$_2$ 
is a layered vdW material. 
The interlayer coupling does not significantly 
alter the electronic structures. 
Nevertheless, a shape peak appears 
at the Fermi level $E_F$ in the monolayer case as shown in Fig.~2(a).
The DOS at $E_F$ is about $N(E_F) \approx$ 6.4 state$/$eV,
much higher than the bulk value. 
Such high DOS 
suggests the presence of VHS on the band structures.
We thus plot
the band structures along high symmetry $k$-path in Fig.~2(b) 
and observe that
only the $d_{z^2}$ band crosses the Fermi level.
On the $\Gamma -M$ path there is a maximum at $\Gamma$ 
and on the $\Gamma - K$ path a minimum at $S$ appears. 
These characteristics indicate the presence of a saddle point. 
To show more details, 
we plot this band on the whole BZ in Fig.~2(c),
where there are six saddle points on the $\Gamma - K$
and $\Gamma - K^\prime$ paths. 
Figure~2(d) zooms in the band structures around
the saddle point $S$.
The red and blue solid lines are the band dispersions along $\Gamma - K$
and its orthogonal directions, which represent
hole-like and electron-like dispersions,
respectively.
These dispersions
evidently show the topology of a saddle point~\cite{PhysRevB.92.085423}.
Therefore, we have demonstrated that the high DOS and its divergent behavior
are due to the presence of saddle-points VHS
on the band structures.

\begin{figure}
\includegraphics[width=0.48\textwidth]{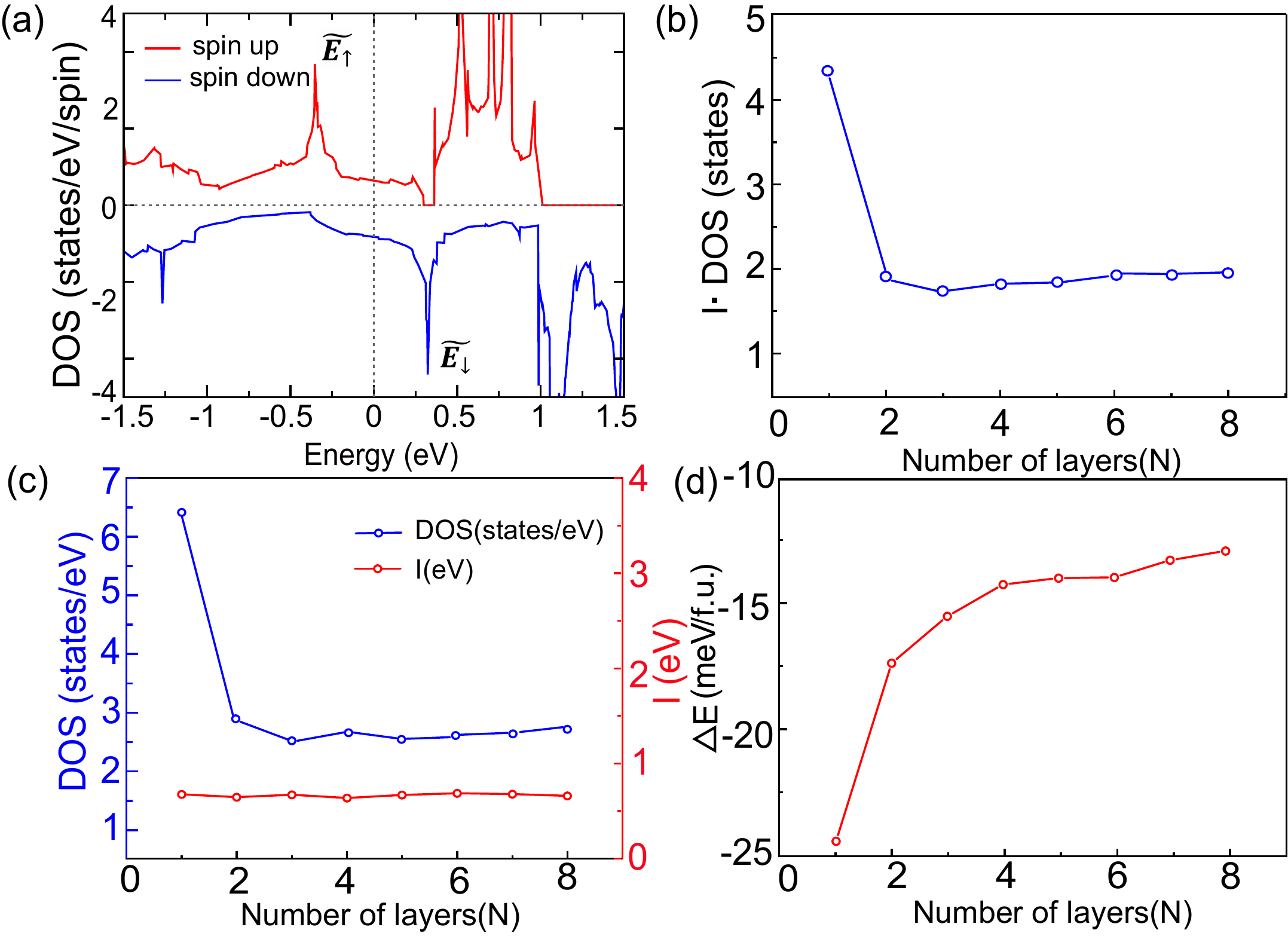}
\caption{ 
    Results of the monolayer 1T-VSe$_2$ in the FM phase.
    (a) The spin polarized DOS. 
 $\tilde{E}_{\uparrow}$ and  $\tilde{E}_{\downarrow}$
denote the energy shift of the VHS peaks, which are used 
to evaluate the energy difference between the spin-up and spin-down states. 
    (b) The evolution of the Stoner criterion with respect to the number of layers $N$.
    (c) The evolutions of DOS at the Fermi level $N(E_F)$ and the Stoner parameter $I$ with respect to $N$.
    (d) The energy difference $\Delta E = E_{FM} - E_{NM}$ as a function of $N$.
}
\label{fig3}
\end{figure}
\noindent

We suggest
that the VHS in monolayer 1T-VSe$_2$ may cause FM instability
according to the phenomenological Stoner theory~\cite{stoner1938collective},
which states that the FM phase is favored
when the Stoner criteria
$N(\epsilon_f) \cdot I > 1$ is satisfied.
Here $N(\epsilon_f)$ is the DOS at the Fermi level in the NM state,
and $I$ is the Stoner parameter that measures
the strength of the magnetic exchange interaction,
which is related to the energy splitting between the spin-up and spin-down states
in the FM phase via the following formulas~\cite{blundell2003magnetism}
\begin{align}
\begin{split}
E_{\uparrow} (k) & = E_0(k) - I \frac{n_{\uparrow}}{n} , \\
E_{\downarrow} (k) & = E_0(k) + I \frac{n_{\downarrow}}{n}. \\
\end{split}
\end{align}
Here $E_0(k)$ is the energy
of the NM phase,
$E_{\sigma}(k)$ and $n_{\sigma}$
are the energy and number of electrons with spin
$\sigma$~$(\sigma = \uparrow,\downarrow)$ in the FM phase, respectively.
The total number of electrons
is $n = n_{\uparrow} + n_{\downarrow}$.
Since only the $d_{z^2}$ band is responsible for the Stoner instability in monolayer 1T-VSe$_2$,
$n_{\uparrow}$ and $n_{\downarrow}$
can be estimated as 1.
Therefore we have $n = 2$.
Finally, the Stoner parameter $I$ can be estimated as
$E_{\downarrow}(k) - E_\uparrow(k)$.

Figure~3(a) presents the DOS of the spin-up and 
spin-down states in the FM phase
of the monolayer 1T-VSe$_2$. 
By comparing with the NM results in Fig.~2(a), 
we observe that the sharp VHS peak
splits into
two peaks, 
which is a typical signature of the FM exchange 
interaction. We assume the exchange interaction 
is $k$-independent and use the energy difference
of the VHS peaks to evaluate its average 
magnitude~\cite{blundell2003magnetism}, 
which gives $I = 0.68$~eV. 
Together with $N(E_F) = 6.4$~state/eV in the NM phase, 
we obtain a Stoner criterion $N(E_F) \cdot I = 4.3 $. 
This large value indicates 
a strong FM instability 
in the monolayer 1T-VSe$_2$.

Recent experiments have shown
that the monolayer 1T-VSe$_2$ exhibits FM order, while 
the bulk 1T-VSe$_2$ displays a NM property~\cite{bonilla2018strong,C3TC31233J,
doi:10.1002/adma.201903779,
doi:10.1021/acsnano.9b02996,
PhysRevB.101.035404}. 
To study this transition, 
we show 
the evolution of
the Stoner criterion
$N(E_F)\cdot I$
with respect to the number of layers $N$ in Fig.~3(b), 
from which a drastic decrease of $N(E_F)\cdot I$
from the mono- to the bilayer is observed,
indicating the decrease of FM instability
in the bilayer 1T-VSe$_2$. 
We further plot the evolutions of
$N(E_F)$
and $I$ with respect to $N$ in Fig.~3(c),
which clearly shows that
the drastic decrease is due to the decrease
of $N(E_F)$ since
$I$ is insensitive to $N$ [see the red curve in Fig.~3(c)].
We notice that when $N \ge 2$, the Stoner criterion $N(E_F)\cdot I$, 
density $N(E_F)$, and Stoner parameter $I$
fluctuate slightly around their saturated values. 
These results prove that when the system transforms 
from the monolayer to bulk, only the monolayer exhibits 
a strong FM instability. 
Our result well explains the recent experiment by Bonilla~\emph{et al.}, where 
a strong FM signal has been detected in the monolayer 
while the bilayer has a significantly weak FM signal comparable with the bulk~\cite{bonilla2018strong}. 
This trend 
is also manifested by the energy difference $\Delta E = E_{FM} - E_{NM}$
between the FM and NM phases as depicted in Fig.~3(d), 
from which we observe that 
the maximal energy difference occurs in the monolayer case,  
and a drastic decrease of $\Delta E$ takes place 
from the monolayer to the bilayer.


\begin{figure}
\includegraphics[width=0.48\textwidth]{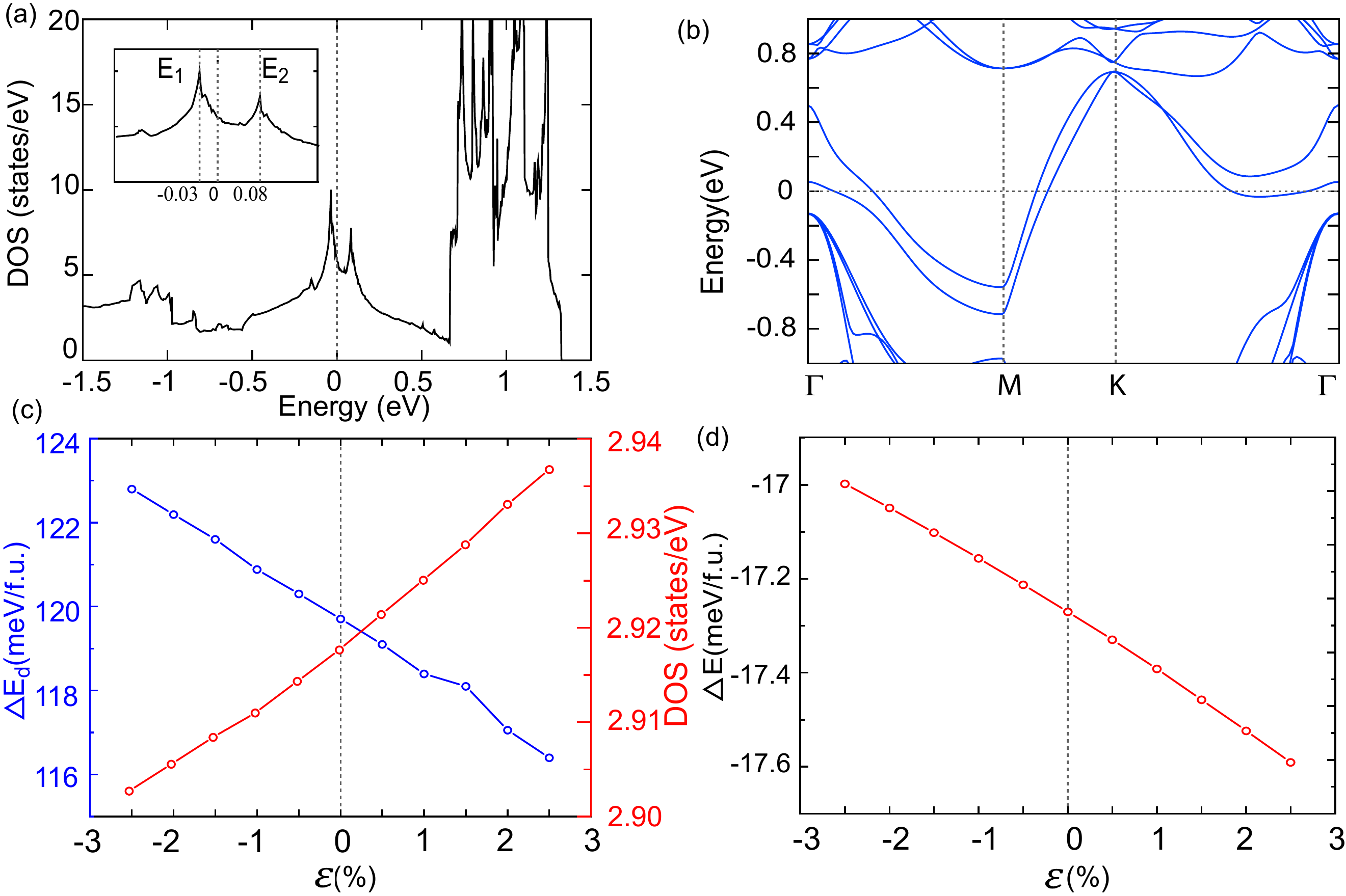}
\caption{ 
    Results of the bilayer 1T-VSe$_2$.
(a) and (b) are the NM DOS and band structures, respectively. 
 The inset of (a) shows the detail of the two peaks.
(c) The blue curve shows the evolution of the  
    two peaks' energy splitting $\Delta E_d$ with the ratio of interlayer distance $\varepsilon$. 
    The red curve shows the $N(E_F)$ 
    as a function of $\varepsilon$. 
(d) shows the energy difference $\Delta E=E_{FM} - E_{NM}$ as a function of $\varepsilon$.
}
\label{fig4}
\end{figure}

To understand the drastic decrease of $N(E_F)$
from the mono to the bilayer,
we plot the NM total DOS of bilayer 1T-VSe$_2$
in Fig.~4(a). 
Due to the vdW nature, 
the bilayer DOS 
is very similar 
with that of the monolayer, 
except that 
two peaks emerge near the Fermi level 
at $E_1 = -0.03$ 
and $E_2 = 0.08$~eV
[see the inset in Fig.~4(a)]. 
These two peaks also originate from the saddle points
on the band structures as shown in Fig.~4(b). 
We notice 
that two bands are 
crossing the Fermi level. 
They are contributed by the $d_{z^2}$ orbitals of 
the two V atoms in the bilayer unit cell. 
The upper and lower $d_{z^2}$ bands 
are anti-bonding and bonding states, respectively. 
The energy difference of these two $d_{z^2}$ bands 
at the $\Gamma$ is about 0.44~eV, 
which gives an estimation of the interlayer coupling strength. 
The VHS splitting at the $S$ point is determined by the two minima 
on the $\Gamma - K$ path, which is about 0.11~eV, consistent with 
the two peaks on the DOS in Fig.~4(a). 
Such splitting is larger than that in typical vdW 
materials~\cite{ersan2019exploring,
jung2018rigorous,
mcguire2015coupling}. 
This is because that the VHS peak in 1T-VSe$_2$ is mainly contributed by 
the $d_{z^2}$ orbitals, 
whose lobes from interlayer V atoms are head-to-head aligned along the 
$z$-direction and form relatively strong $dd\sigma$ bonds. 
As a result, the VHSs no longer present at the Fermi level, 
and the $N(E_F)$ is significantly reduced. 
Finally, the Stoner criterion $N(E_F)\cdot I$ decreases severely, 
and the FM instability is weakened.



To summarize, 
the transition from 
the bulk NM phase 
to the monolayer FM phase 
in 1T-VSe$_2$ can be understood as follows. 
In the bulk system, 
the coupling of $d_{z^2}$ orbital between interlayer V atoms 
splits the VHSs away from the Fermi level. 
Especially from the bilayer to the monolayer, 
the enhanced confinement effect 
eliminates the interlayer $d_{z^2}$ coupling. 
Thus the VHSs are pushed to the Fermi level, 
which leads to a drastic enhancement of the $N(E_F)$ 
and the Stoner criterion $N(E_F) \cdot I$.  
Eventually, this enhanced $N(E_F)$ causes a strong FM instability in the monolayer. 
In other words, 
the confinement effect
in the monolayer 1T-VSe$_2$
prevents the interlayer coupling
between the $d_{z^2}$ orbitals of V atom
and pushes the saddle-point VHS at the Fermi level,
which results in a large Stoner criterion 
and leads to a stable FM ground state.
Our numerical results can well explain recent experiments~\cite{bonilla2018strong,
C3TC31233J,
doi:10.1002/adma.201903779,
doi:10.1021/acsnano.9b02996,
PhysRevB.101.035404}.

Based on this understanding, 
we expect 
that the magnetic property of a few-layers 1T-VSe$_2$ 
can be tuned by the interlayer distance $d$~\cite{zhou2012tensile}
[see Fig.~1(a)].
The evolution
of the VHS splitting $\Delta E_{d} = E_2 - E_1$ between the two peaks 
in the bilayer 1T-VSe$_2$ 
with the ratio $\varepsilon = (d-d_0)/d_0$
is shown in Fig.~4(c). 
Here $d_0 = 3.067$~{\AA} is the interlayer distance of the bulk. 
The corresponding evolution of $N(E_F)$ with $\varepsilon$ is also shown in this figure. 
We observe 
that $\Delta E_d$ monotonically decreases with the increase of $d$. 
This means the $dd\sigma$ bond is weakened 
when the lobes of $d_{z^2}$ orbitals in adjacent layers are moving apart. 
During this process, the $N(E_F)$ monotonically goes up, 
reflecting an enhanced Stoner FM instability. 
The enhancement of the Stoner instability 
is also confirmed 
by the energy differences 
$\Delta E$ between the FM and NM phases in Fig.~4(d).
We find that $\Delta E$ also monotonically decreases with $\varepsilon$, 
which indicates that 
the FM phase becomes more and more stable when $d$ increases. 
This effect provides a useful 
route to control
the NM to FM transition in few-layers 1T-VSe$_2$ 
through enlarging the interlayer distance $d$. 
It also sheds light on tuning the Curie temperature 
of 1T-VSe$_2$ through nanoengineering. 
Further experimental 
studies are highly desirable to 
verify these conjectures.


\section{ Acknowlegments } 
This work is supported by the Ministry of Science and Technology of China (No. 2018YFA0307000) 
and the National Natural Science Foundation of China (No. 11874022).

\bibliography{ref}

\end{document}